\documentclass[prb,
twocolumn,superscriptaddress,amsmath,amssymb]{revtex4}
\usepackage{xcolor}
\usepackage{graphicx}
\usepackage{hyperref}
\definecolor{JN-color}{rgb}{1,0,0}

\newcommand{\vek}[1]{\boldsymbol{#1}}
\newcommand{\unitvek}[1]{\hat{\boldsymbol{#1}}}

\newcommand{\doo}{\partial}
\newcommand{\im}{i}

\newcommand{\unitvec}[1]{\hat{\vek{#1}}}

\newcommand{\id}{\boldsymbol{1}}

\begin{document}

\title{Anomalous chiral transport with vorticity and torsion: Cancellation of two mixed gravitational anomaly currents in rotating chiral $p+ip$ Weyl condensates}

\author{J. Nissinen}
\affiliation{Low Temperature Laboratory, Department of Applied Physics, Aalto University, FI-00076 AALTO, Finland}

\author{G.E.~Volovik}
\affiliation{Low Temperature Laboratory, Department of Applied Physics, Aalto University, FI-00076 AALTO, Finland}
\affiliation{Landau Institute for Theoretical Physics, acad. Semyonov av., 1a, 142432,
Chernogolovka, Russia}

\date{\today}

\begin{abstract}
Relativistic gravitational anomalies lead to anomalous transport coefficients that can be activated at finite temperature in hydrodynamic and condensed matter systems with gapless, linearly dispersing fermions. One is the chiral vortical effect (CVE), an anomalous chiral current along the system's rotation axis, expressed in terms of a gravimagnetic field in a rotating frame and a mixed gravitational anomaly. Another one arises in the presence of hydrodynamically independent frame fields (and spin-connection) and leads to the thermal chiral torsional effect (CTE). We discuss the relation of CVE, CTE and gravitational anomalies for relativistic fermions from the perspective of non-zero torsion and the Nieh-Yan anomaly. The DC transport coefficient induced by the two gravitational anomalies are found to be closely related and equal. At level of linear response, their difference is demarcated whether or not torsion is non-zero and the existence of non-metric degrees of freedom in the hydrodynamic constitutive relations (with sources). In particular, the relativistic anomaly from torsion is well defined, since instead of a removable ultraviolet (UV) divergent term, the (usually IR) chemical potential or temperature scales enter. This is closely related to the CVE from the 4th order $R^2$-gravitational anomaly and its appearance already in linear response. At the same time, the torsional anomaly is 2nd order in gradients and directly contributes in linear response for CTE, implying also the same for CVE. We clarify this and the hydrodynamic relations leading to the torsional contribution when the currents depend on the frame fields and connection instead of the metric. As an example where the two anomalies are sourced independently, we consider chiral $p+ip$ Weyl superfluids and superconductors at finite temperature. At low-energies in the linear approximation, the system is effectively relativistic along a special anisotropy axis. The hydrodynamics is governed by two velocities, normal velocity $\vek{v}_n$ and superfluid velocity $\vek{v}_s$. The existence of the two thermal anomalies in the condensate follows from the normal component rotation and the dependence of the momentum density on the superfluid velocity (order parameter). In the CVE the chiral current is produced by the solid body rotation of the normal component with (angular) velocity $\vek{v}_n= \boldsymbol\Omega \times \vek{r}$. In the CTE, a chiral current is produced by the vorticity of the superfluid velocity $\nabla \times \vek{v}_s$, which in the low-energy quasirelativistic effective theory plays the role of gravitational torsion from the order parameter. In thermal equilibrium, $\langle\langle \nabla \times  \vek{v_s}\rangle\rangle= 2\boldsymbol\Omega$ on average and the two gravitational anomaly currents cancel each other. This is an example of the Bloch theorem for axial currents, prohibiting finite current in equilibrium, now satisfied as the cancellation of two gravitational anomalies with independent sources: gravimagnetic rotation field and effective condensate torsion from $\vek{v}_s$.  Although the latter represents physically the superfluid vorticity, similar to the CVE, in the low-energy quasirelativistic theory it arises from effective torsion for the normal component chiral fermions.
\end{abstract}
\pacs{
}

\maketitle

 \newpage

\section{Introduction}

Consequences and implications of field theoretic quantum anomalies \cite{Adler1969, BellJackiw1969, Adler2005, EguchiFreund76, AlvarezGaumeWitten84, AlvarezGaumeGinsparg85} in the low-energy hydrodynamic transport in chiral media have recently attracted considerable interest with applications spanning ultrarelativistic quark-gluon plasmas in heavy-ion collisions \cite{FukushimaEtAl08, KharzeevSon11, RenEtAl12, LectNotes13, KharzeevEtAl16}, holographic phenomenology \cite{ErdmengerEtAl09, BanerjeeEtAl11, Landsteiner2011, Landsteiner2015} as well as topological materials with protected quasirelativistic chiral fermions including semimetals, superfluids and superconductors \cite{Bevan1997, KrusiusEtAl98, Volovik2003, SonYamamoto12, KharzeevYee13, ZyuzinBurkov12, Stone12, Zaheed2012, BasarKharzeevZahed2013, SonSpivak13, Landsteiner14, CanLaskinWiegmann14, ChernodubEtAl14, Burkov15, SpivakAndreev16, SumiyoshiFujimoto16, QiKharzeevEtAl16, Lucas2016, Landsteiner2016,GoothEtAl17, Wiegmann18, GaoEtAl21, Chernodub2021, RamshawEtAl21}. 

The quantum field theoretic anomalies are in four dimensions defined by triangle diagrams with gauge and gravitational vertices and in the low-energy hydrodynamic regime lead to the chiral magnetic (CME) and chiral vortical effects (CVE) at non-zero chemical potential and temperature \cite{Vilenkin79, Vilenkin80, FukushimaEtAl08, LectNotes13}: The CME is a chiral current along magnetic field at finite chemical potential(s), while the CVE involves a chiral current induced by hydrodynamic fluid flow with non-zero vorticity, e.g. the angular momentum of a rotating chiral system. Intuitively speaking, the CME is based on the analogy of (chiral) chemical potential with (chiral) temporal gauge fields, whereas the CVE rests in addition to the analogy of rotating frames to (tidal) gravitational forces. 

A physically transparent way to find such transport coefficients is the requirement of positive entropy current  in the presence of anomalous sources, in accordance with the second law of thermodynamics \cite{SonSurowka09}. On the other hand, resting on the analogy with gravity, the appearance of the static, equilibrium anomaly transport coefficients can be organized as an order-by-order derivative expansion of the hydrodynamic conservation laws on non-trivial geometric backgrounds with sources \cite{NeimanOz11, BanerjeeEtAl12, Jensen12, JensenEtAl13}. The gravitational (i.e. thermal \cite{Luttinger64, Chernodub2021}) CVE anomaly current at linear order in response has been related to the gravitational anomalies \cite{Landsteiner2011, LoganayagamSurowka12, JensenEtAl13} that, at least naively, appear only at higher-orders in the hydrodynamic gradient expansion. The precise way how higher-order anomalies can contribute to the transport has lead to many useful insights about relativistic and non-relativistic hydrodynamics, e.g. \cite{SadofyevEtAl11,Golkar2015,HaehlEtAl15,Jain16,Glorioso2017,StoneKim18,FlachiFukushima18,ProkhorovEtAl20,MitkinZakharov2021}.

In a more recent development, there has been renewed interest in the field theoretic and hydrodynamic anomalies with non-zero torsion (and relatedly, so-called pseudo gauge fields) in systems with independent frame fields and connection, in contrast to the metric. Several authors have considered non-zero torsional transport coefficients, given some assumptions beyond standard relativistic field theory and/or finite chemical potential and temperature \cite{SunWan14, HughesEtAl13, ParrikarEtAl14, AbanovGromov14, GromovAbanov15, Zubkov2018, HuangEtAl19, FerreirosEtAl19, Imaki2019, Nissinen20, Copetti20, Laurila20, HuangEtAl20, NissinenVolovik2019, NissinenVolovik2020, LiangOjanen20, Imaki2020, FerreirosLandsteiner20, ManesEtAl21, Liu21}. The independent gravitational anomaly term from torsion \cite{NiehYan1982a, NiehYan1982b, Nieh2007}, the Nieh-Yan anomaly term \cite{Yajima96, ObukhovEtAl97, ChandiaZanelli97, Soo99, PeetersWaldron99, Comments}, is special in that it's 2nd order in the gradients, in contrast to the 4th mixed $R^2$ gravitational anomaly from topological anomaly polynomials, and appears with a dimensionful scale parameter that is, apparently, non-universal and unquantized. With considerable past and more recent history, many authors have argued for and against this term with varying and contrasting results, especially whether a dimensionful ultraviolet (UV) scale can appear in the anomaly coefficient. Here we are contend to remark that this term is bound to vanish in relativistic systems, since there is a counter term available that breaks no additional symmetries, while apriori such a term is possible in non-relativistic systems with a cutoff scale to low-energy chiral transport and fermions. See the experimental identification of this chiral anomaly \cite{Bevan1997, Volovik2003} and how it can be matched consistently with the torsional anomaly both at zero and finite temperatures \cite{Nissinen20, NissinenVolovik2020}. 

Notwithstanding, both the relativistic and non-relativistic torsional anomalies can appear with the (usually) infrared (IR) temperature scale or chemical potentials, similarly to the gravitational contribution to the CVE and we focus here on these terms. Moreover, such a finite temperature (chemical potential) term can be ``universal" as it rest on the anomaly related quantum statistical properties, with the same caveats as the hydrodynamic CVE (CME). In this paper we discuss the chiral torsional effect (CTE) from the perspective of the analogy and similarity with CVE in relativistic systems with chiral fermions. We find that the coefficient is non-zero and directly related to that of the CVE (at the level of linear response), and, in accordance with the previous results found with e.g. the simple Landau level approach to chiral torsional anomalies. In particular, recent papers \cite{FerreirosLandsteiner20, Chernodub2021} finding no torsional contribution can be easily incorporated with our framework by noting that when computing the linear response around flat space, torsion has been actually set to zero with only non-zero metric variations. This is equivalent to whether or not frame fields and torsion appear in the hydrodynamics and needs to be separately addressed at the level of the constitutive relations in the presence of sources. Nevertheless, we find non-zero transport coefficients in the case torsionful sources do appear.

Although a relativistic system (or gravity \cite{BahamondeEtAl21}) with torsion remains to be identified in the real world, we note that geometric torsion has been found to be inherent in the hydrodynamics of many condensed matter systems, including elasticity, see e.g. \cite{KondoBilbyKroner, DzyaloshinskiiVolovick84, Kleinert, BeekmanEtAl17, Boehmer20}, topological quantum Hall and paired systems \cite{BradlynRead15, GolanStern18, Nissinen20}, semimetals \cite{SunWan14, HughesEtAl13, ParrikarEtAl14, HuangEtAl19, Laurila20}, crystalline insulators \cite{Gromov19, NissinenVolovik19b} and in general non-relativistic Newton-Cartan spacetimes \cite{Son13, GeracieEtAl17, HansenEtAl20}. As an example of the difference and similarities of the two independent anomaly sources, we discuss the two anomalies in a rotating chiral (non-relativistic) $p+ip$ Weyl superfluid or superconductor with a normal component and vortex lattice with vorticity and low-energy effective torsion, respectively. This gapless system is not strictly relativistic \cite{Horava2005} but instead, at low-energies, described by spatially anisotropic Newton-Cartan geometry that identifies the origin of the effective torsionful quasiparticle geometry in terms of the more directly familiar independent hydrodynamic variables and symmetries. However, in the linear approximation, the low-energy theory satisfies effective (local) Lorentz invariance along the special anisotropy direction, therefore the ingredients for (relativistic) CVE and CTE can be applied if the rotation, vorticity and torsion are along this special axis.  

The rest of this paper is organized as follows. In Sec. \ref{sec:intro} we review the CVE and chiral anomaly. In Sec. \ref{sec:gravitational} we discuss the gravitational anomaly perspective of CVE and CTE. Sec. \ref{sec:Kubo} presents the Kubo formula argument for CTE and its relation to CVE. Sec. \ref{sec:cancellation} discusses the cancellation of CVE and CTE in rotating chiral $p+ip$ Weyl condensates. Conclusions and Outlook end the paper with an Appendix containing geometric formulas.

\section{Chiral anomaly, chiral magnetic and vortical effects}\label{sec:intro}
To set the notation and introduce the anomalous currents, we consider the low-energy, effective hydrodynamic theory relativistic Dirac fermions in a chemical potential of a low-energy hydrodynamic fluid, $\gamma^0\mu \to \gamma^{\mu} \mu u_{\mu}$ \cite{SadofyevEtAl11},
\begin{align}
S_{\rm hydro}[\psi,\psi^\dagger] = \int d^4x~ \overline{\psi} \gamma^\mu \im(\doo_{\mu} -iqA_{\mu}) +\mu \overline{\psi}\gamma^\mu u_{\mu}\psi. \label{eq:CME}
\end{align}
Here $u_{\mu}$ is a local velocity field for a fluid element in its rest frame, $u^\mu u_{\mu} = 1$, a generalized chemical potential in the low-energy effective hydrodynamical theory. As is obvious, $u_{\mu}$ enters similarly as a U(1) gauge field, 
\begin{align} 
\frac{\delta S}{\delta u_{\mu}} = \mu \overline{\psi}\gamma^{\mu}\psi = \mu J^\mu .  \label{eq:u_variation}
\end{align}
Relativistically, the $u_{\mu}$ couples to  $\mu J^\mu$ with units of (energy-)momentum. The fermions are massless and cannot equilibrate with the low-energy fluid velocity $u_\mu$: The presence of the low-energy background fluid $u_{\mu}$ singles out a preferred frame. Similarly, one can introduce an axial chemical potential $\mu_5\gamma^5$ velocity field $\mu_5 u_{5\mu} \gamma^5$. 

For now, we set $\mu=\mu_5 = \mu_\chi/2$ and focus on single right-handed fermion $\chi=+1$ with a global (and local) U(1). Calculating the anomalous current of \eqref{eq:CME}, loosely via the implied substitution $qA_\mu \to qA_{\mu} + \mu u_{\mu}$, we arrive to
\begin{align}
J^\mu_\chi = \frac{\epsilon^{\mu\nu\lambda\rho} }{8\pi^2}(eA_{\nu} + \mu u_{\nu}) (e\doo_{\lambda}A_{\rho} +\mu\doo_{\lambda}u_{\rho}).
\end{align}
The gauge-invariant cross terms give the CME, e.g. in the simplest case when $u_{\mu}=(1,0,0,0)$. Setting $A_{\mu} =0$, we obtain the chiral vortical effect (at $T=0$ and to lowest order in $\mu$)
\begin{align}
J^{\mu}_\chi = c_\chi\frac{\mu_{\chi}^2}{8\pi^2} \epsilon^{\mu\nu\lambda\rho}u_{\nu}\doo_{\lambda}u_{\rho},
\end{align}
where $c_{\chi} = +1$ is the coefficient of the chiral U(1) anomaly for right-handed fermions. 

\section{Mixed gravitational and torsional anomalies}\label{sec:gravitational}

On the other hand, we can consider gravitational action of Dirac fermions, understood as an effective action in the hydrodynamic regime around a near-equilibrium background with sources,
\begin{align}
S_{\rm grav}[\psi,\overline{\psi}] = \int d^4x e~ \overline{\psi} e_a^{\mu}\gamma^a \im(\doo_{\mu}-\frac{\im}{2}\omega_{\mu ab} \gamma^{ab})\psi 
\end{align}
where $g_{\mu\nu} =  e^a_{\mu} e^b_{\nu}\eta^{ab}$ in terms of the tetrad, $e^\mu_a$ the inverse and $\omega_{\mu ab}\gamma^{ab}$, $\gamma^{ab} = \frac{\im}{4}[\gamma^a,\gamma^b]$, is the spin-connection in the spin $1/2 \oplus 1/2$ Weyl-representation. We consider two cases here: when $S_{\rm eff} = S_{\rm grav}[g_{\mu\nu}]$ depends only on the metric $g_{\mu\nu}$ and when $S_{\rm eff} = S_{\rm grav}[e^a,\hat{\omega}_{\mu}]$ depends on $e^a_{\mu}$ and $\hat{\omega}_{\mu}$ independently. To compare with $S_{\rm grav}$ with $S_{\rm hydro}$, we set $g_{\mu\nu} = \eta_{\mu\nu} + h_{\mu\nu}$, where $h_{\mu\nu}$ is small perturbation, 
\begin{align}
ds^2 = g_{\mu\nu}dx^\mu dx^{\nu} \nonumber
= dt^2-2\vek{u}\cdot d\vek{x} dt - d\vek{x}^2 
\end{align}
and $-u_i = h_{ti}$ is a small velocity. In terms of tetrads/vierbein $g_{\mu\nu} =\eta_{ab} e^{a}_{\mu}e^b_{\nu}$,
\begin{align}
e^0_{\mu} &= (1,-\vek{u}),\quad e^{m}_{\mu} = \delta^m_{\mu} \nonumber\\
e^{\mu}_0 &= (1,0),\quad  e^\mu_m = (u_m,\delta^\mu_m) \label{eq:tetrad_background}. 
\end{align}
The torsion free spin-connection corresponding to this variation is collected in the Appendix. With non-zero torsion, the connection is different. In terms of variations \cite{GeracieEtAl17}
\begin{align}
\delta S &= \int d^4x~ e  [\tilde{T}_a^{\ \mu}\delta e^a_{\mu} + s_{\ ab}^{\mu}\delta \omega^{ab}_{\mu}] \nonumber\\
&= \int d^4x~e [ T_a^{\ \mu}\delta e^a_{\mu}  + s^{\mu}_{\ ab} \delta K_{\mu}^{\ ab}] \label{eq:torsion_variations} \\
&= \int d^4x~e [ T_a^{\ \mu}\delta e^a_{\mu} + S_{a}^{\ \mu\nu} \delta T^a_{\mu\nu}] \nonumber
\end{align}
where 
\begin{align}
\tilde{T}_a^{\ \mu} = \frac{1}{e} \frac{\delta S}{\delta e^b_{\nu}}, \quad s_{\ ab}^{\mu} &= \frac{1}{e} \frac{\delta S}{\delta \omega^{ab}_{\mu}}, \label{eq:tetrad_variations}
\end{align}
are the tetrad energy-momentum and (intrinsic) spin currents. The $K^a_{\mu b} = (\omega - \mathring{\omega})^a_{\mu b}$ is the contorsion tensor, $\mathring{\omega}_{\mu}$ the Christoffel connection fully determined by $e^a_{\mu}$ and $S_{a}^{\ \mu\nu} = \frac{1}{2}\eta_{ab}e^b_\lambda (s^{\mu\nu\lambda}-s^{\nu\lambda\mu}-s^{\lambda\mu\nu})$. The $\tilde{T}^a_{\ \mu}$ and $T^a_{\ \mu}$ differ by torsion and spin-current terms \cite{GeracieEtAl17}. The different variations arise when either $e^a$ and $\omega^{ab}$ are treated independent, the $e^a$ and $K^{ab}$ or, finally, $e^a$ and $T^a$. We shall see blelow that around flat space, the different variations of \eqref{eq:torsion_variations} are simply related.

These variations replace the metric variation with fixed connection torsion free connection
\begin{align}
\delta S = \int d^4x ~e \frac{1}{2}T^{\mu\nu}\delta g_{\mu\nu} \label{eq:metric_variation}
\end{align}
where $T^{\mu\nu} = \frac{1}{2}( e_a^\mu  T^{a\nu} + e^\nu_a T^{a\mu} )$ is the symmetric energy-momentum tensor and $\delta g_{\mu\nu} = e^{a\mu}\delta e^\nu_a + e^{a\nu}\delta e^{\mu}_a$ the symmetric variation. We have assumed that the Lorentz-anomaly vanishes and $T^a_{\mu}$ can be made symmetric, see Sec. \ref{sec:Conclusions} for discussions.

\subsection{Gravitational anomaly}
The hydrodynamic effective action $S_{\rm grav}$ contains only gravitational (geometric) fields. Now, to the lowest order in the absence of torsion, the mixed chiral-gravitational anomaly is given as
\begin{equation}
 \nabla_\mu J_{\chi}^\mu = \frac{d_{\chi}}{768\pi^2}  e^{\mu\nu\lambda\rho} R^\alpha_{\beta\mu\nu} R^\beta_{\alpha\lambda\rho} \,.
\label{a2}
\end{equation}
For single right-handed chiral fermion with global U(1), $d_{\chi}=1$. This leads to the chiral vortical effect:
\begin{equation}
J^\mu_{\chi} =  d_{\chi}\frac{T^2}{24} \omega^\mu,\quad \omega^\mu = \epsilon^{\mu\nu\lambda\rho} u_\nu \partial_\lambda u_\rho\,\,  \,.
\label{Jmu}
\end{equation}
The chiral vortical effect and mixed gravitational anomaly have the same coefficient, which follows from free theory and holographic examples, see e.g. Ref. \cite{Glorioso2017,Golkar2015,Landsteiner2011,Landsteiner2011b} and references therein. Above, we discussed this anomalous current induced by chemical potential from the chiral anomaly with coefficient $c_{\chi}\mu^2$, below the $\propto T^2$ term with coefficient $d_\chi$ is discussed via the response to $\delta g_{ti}$ around flat space and its relation to the torsional anomaly. The connection is reviewed below at level of linear response Kubo formula for free fermions.

\subsection{Chiral torsional effect and Nieh-Yan gravitational anomaly}

There is also the chiral torsional effect (CTE) \cite{Zubkov2018,Imaki2019} from the mixed chiral-torsional anomaly (thermal Nieh-Yan anomaly \cite{NissinenVolovik2020,NissinenVolovik2019, HuangEtAl20, Imaki2020}):
\begin{align}
J_5^\mu &=-t_\chi \frac{T^2}{48}\epsilon^{\mu\nu\lambda\rho} e^0_{\nu}T^0_{\lambda\rho}, 
\label{j5}
\end{align}
Here $t_{\chi}=d_{\chi}=1$ for a right-handed fermion, as is discussed below. In linear response, $J^k_{5} = t_{\chi} \frac{T^2}{48} \epsilon^{kij}T^0_{ij}$. The connection to CVE follows in the simplest terms via the identification $u_\mu = u_{a}e^a_{\mu} = e^0_\mu$, where $u_a = (1,0,0,0)$. Whence the CVE becomes
\begin{align}
\epsilon^{\mu\nu\lambda\rho}u_{\nu}\doo_{\lambda}u_\rho = \epsilon^{\mu\nu\lambda\rho}u_a u_b e^a_{\nu}\doo_{\lambda}e^b_\rho = \epsilon^{\mu\nu\lambda\rho}e^0_{\nu}\doo_{\lambda}e^0_{\rho}.
\end{align}
Relativistically $u_\mu u^{\mu} = u^a u_a = 1$ and in the fluid rest frame $u_a = (1,0,0,0)$. We note that non-relativistically, one usually fixes the time-like one form $u_a = (1,0,0,0)$ as the Newtonian clock form, whereas the velocity $u^a=(1,\vek{u})$ is independent with spatial vorticity $\nabla \times \vek{u}$.

\section{Torsional Kubo formula and currents} \label{sec:Kubo}
\subsection{Kubo formula with torsion}
Building on the earlier and recent work, let us present the Kubo formula for the CVE and derive from thereon its version with non-zero torsion. The CVE current is
\begin{align}
J_\chi^{\mu} = \sigma^{V}_{\chi} \epsilon^{\mu\nu\lambda\rho} u_{\nu}\doo_{\lambda}u_{\rho}
\end{align}
where $\sigma^V_{\chi}$ is the chiral vortical conductivity. For applications, we are interested in the axial current $J_{5} = \sum_{\chi} \chi J_{\chi} = J_{+} -J_{-}$. Since $u_{\lambda} = h_{t\lambda} = (1, u_i)$, its Kubo formula reads (no sum over $m$)
\begin{align}
\sigma_{\chi}^V = \lim_{k\to 0} \frac{\im \epsilon_{ijm}}{2 k^m} \langle J_{\chi}^i T^{tj} \rangle(k)_{\omega=0}
\end{align}
where $\delta S = \int d^4x\sqrt{-g} T^{t\lambda}h_{t\lambda}$. This is evaluated as
\begin{align}
\sigma_{\chi}^V =& \lim_{k\to 0} \frac{\im \epsilon_{ijm}}{2k^m}\\ 
&\times \int d^4k~ e^{\im k\cdot (x-x')} \theta(t-t') \langle [ J_{\chi}^i(x), T^{tj}(x') ]\rangle_{\omega=0}. \nonumber
\end{align}
This correllator was computed in \cite{Landsteiner2011} for free fermions. It is composed of two terms, $\sigma^V_{\chi} = \sigma^{\tilde{T}}_{(0j)} + \sigma^{\tilde{T}}_{j0}$, where we defined the torsional conductivity $\sigma^{\tilde{T}}_{(a\mu)} = \lim_{k\to 0}\im\epsilon_{ijm}(k^m)^{-1}\langle [J^i_{\chi}, \tilde{T}^{a\mu}]\rangle_{\omega=0}$. This follows directly from
\begin{align}
T^{tj} &= \frac{i}{2} \overline{\psi}( \gamma^0 \doo^{j} + \gamma^j \doo^t)\psi \nonumber\\
&= \frac{1}{2} (e^t_0 \tilde{T}^0_{\ i}g^{ij} + e_m^j \tilde{T}^{m}_{\ t}g^{tt} ) \\
&= \frac{1}{2}(\tilde{T}^{0j} + \tilde{T}^{jt}) \nonumber.
\end{align}
We make no distinction between the indices, as in linear response $e^\mu_a = \delta^\mu_a$, $g^{\mu\nu} = \eta^{\mu\nu}$, and
\begin{align}
\tilde{T}^a_{\ \mu} := \frac{1}{e} \frac{\delta S}{\delta e^\mu_a}\bigg\vert_{\hat{\omega}_{\mu}} = - e^b_{\mu} e^{\nu}_a \tilde{T}_{b}^{\ \nu} . 
\end{align}
When torsion vanishes, we should use $T^{0j}$ corresponding to the metric variation $\delta g_{\mu\nu} = h_{\mu\nu}$, while for non-zero torsion we should use $\tilde{T}_a^\mu$ and $\delta e^a_{\mu}$. We now focus explicitly on the latter case with torsion.

The vacuum subtracted, finite result for of both static correlators involving $\tilde{T}^{0j}$ and $\tilde{T}^{jt}$ turns out to be identical \cite{Landsteiner2011, Vilenkin79}, as $k\to 0$ and $n_F(x)$ the Fermi distribution,
\begin{align}
\sigma^{\tilde{T}}_{(0j)} = \sigma^{\tilde{T}}_{(jt)} &= \frac{1}{8\pi^2} \int_0^\infty dq q [n_{F}(q+\mu)+n_{F}(q-\mu)] \nonumber\\
&= \frac{\mu_\chi^2}{8\pi^2} + \frac{T^2}{24} . \label{eq:torsion_Kubo}
\end{align}
From the definitions of $\tilde{T}^{a}_{\ \mu}$ and $\tilde{T}_b^{\ \nu}$, evaluated on the background \eqref{eq:tetrad_background} with non-trivial $e^0_{\mu}$, this translates to
\begin{align}
J_\chi^\mu = \sigma^{\tilde{T}}_{(0i)} \epsilon^{\mu\nu\lambda\rho}e^0_{\nu}\doo_{\lambda}e^0_{\rho} = \frac{\sigma^{\tilde{T}}}{2} \epsilon^{\mu\nu\lambda\rho} e^0_{\nu} T^0_{\lambda\rho}  \label{eq:timelike_torsion}
\end{align}
where $\sigma^{\tilde{T}} = c_\chi \frac{\mu_\chi^2}{8\pi^2} + d_{\chi}\frac{T^{2}}{24}$.

We note that precisely the same distribution integral for $\sigma^{\tilde{T}}_{(a\mu)}$ as in \eqref{eq:torsion_Kubo} was found utilizing the Landau level approach \cite{Laurila20}, there just with non-trivial spatial $e^m_{\mu}$, as applied to finite temperatures \cite{NissinenVolovik2020, HuangEtAl20}. In hindsight, the same conclusion follows of course trivially from the CVE with $e^0_i = u_i$ and the assumption that torsion is non-zero (i.e. $\hat{\omega}_\mu = 0$). This connection was first utilized in \cite{Zubkov2018}. The main result of \cite{FerreirosLandsteiner20} is that
\begin{align}
\sigma^V_{5}(\mu=\mu_5=0) = \textrm{Tr}(T_A)_R - \textrm{Tr}(T_A)_L = d_A
\end{align}
is the gravitational anomaly coefficient for $N_f$ fermions with $T_{A}$ the generators of a global symmetry $G$. For a pair of U(1)$_{\chi}$ Weyl fermions with opposite chirality $\chi=\pm 1$, we recover the axial 
$\sigma^{\tilde{T}}_{5} = \sigma_{5}^V = \frac{\mu^2+\mu_5^2}{4\pi^2} + \frac{T^2}{12}$, see below.

In linear response around flat spacetime, the chiral vortical and torsional responses are simply related with distinction only in the chosen hydrodynamic degrees of freedom (metric vs. tetrad and connection). Nevertheless, the non-zero quantities $\sigma^{\tilde{T}}_{(a\mu)}$ in the correlation function of the Kubo formula show explicitly that on torsional backgrounds, the NY form contributes to the current with the same DC coefficient. If torsion vanishes and $\tilde{T}^{a}_{\ \mu}$ is symmetric, the CVE formula is recovered. This is equivalent to that only $T^{\mu\nu}$ and the metric $g_{\mu\nu}$ enter the response or more generally the hydrodynamic constitutive relations with sources. In contrast, on torsional backgrounds, $\tilde{T}_a^{\ \mu}$ and $e^a_{\mu}$ are the appropriate (hydrodynamic) variables and sources. 

This completes our review of the Kubo formula with (and without) torsion. Next we discuss the connection of these results to the arguments of \cite{FerreirosLandsteiner20}, and in contrast find explicit torsional currents in correspondence with the finite $\sigma^{\tilde{T}}$.

\subsection{Chiral torsional currents}

The above Kubo formula can be used to compute the response to torsion. Following \cite{FerreirosLandsteiner20}, let us do non-zero variations of torsion in two independent ways: i) First varying the connection, holding the spin-connection fixed and then ii) keeping the spin-connection fixed, varying only the tetrad. This variations correspond to the independent variables corresponding to the first two torsionful variations in \eqref{eq:torsion_variations}. Throughout, we will keep all other sources absent, so that the current is zero only if torsion does not contribute.

i) In this case $\delta \omega$ is the only non-zero variation. Since the tetrad is fixed, the spin-connection is then exclusively (con)torsional and contributes only through an axial term from totatally antisymmetric torsion. 

In more detail, the NY form is given as $\delta (e_a \wedge T^a) = e_a \wedge \delta T^a = e_a \wedge \omega^a_{\ b} \wedge e^b$, with $\delta \omega$ contributing as effective $\delta A_5$,  see e.g. \cite{Soo99},
\begin{align}
\nabla_{\mu} = \doo_{\mu} - \im\gamma_5 \frac{1}{8}\epsilon_{\mu}^{\ \nu\rho\lambda} \delta T_{\nu\lambda\rho} \equiv \doo_{\mu}+\im\gamma^5\delta A^T_{5\mu}.\label{eq:torsion}
\end{align}
In particular, for space like antisymmetric torsion, from terms with $\delta A^{T}_{5t}$, where the superscript reminds that this axial field originates from variation of (con)torsion \emph{not} a gauge field or chemical potential. The result is that torsion contributes to the current $\star J_5 = \sigma^{T} e_a \wedge T^a$ with the coefficient:
\begin{align}
J_{5}^t &= \sigma_5^{T} \delta A_{5}^{Tt} \nonumber\\
&=  \frac{1}{8}\sigma_5^T \delta (\epsilon^{tijk} T_{ijk}) \label{eq:torsion2} \\
&=  -\frac{1}{4}\sigma_5^T \star (e_m \wedge \delta T^m)^t \nonumber
\end{align}
where the last line defines the one form dual of the three form $e_a\wedge T^a$, equal to $e_m \wedge T^m$ in linear response. The current is \emph{exclusively} from the torsional NY term. It vanishes if torsion is zero. With the relations \eqref{eq:torsion2} and \eqref{eq:torsion}, the coeffiecient is found easily using 
\begin{align}
\frac{1}{4}\sigma^{T}_5 = \frac{1}{4}\langle J_5^t J_5^t \rangle =\frac{1}{4}\frac{d\rho}{d\mu^T_{5}} = \frac{\mu^2+\mu^2_5}{4\pi^2} + \frac{T^2}{12} . \label{eq:thermal}
\end{align}
equal to \eqref{eq:torsion_Kubo}, now with space-like torsion. In contradistinction with Eqs. \eqref{eq:torsion2} and \eqref{eq:thermal}, the quantity $\frac{d\rho}{d\mu_{5}^T}$ was not included in coefficients $c_{T\perp}$ in Eq. 21 of \cite{FerreirosLandsteiner20}. In addition, the various coefficients in Eqs. 28-33 of Ref. \cite{FerreirosLandsteiner20} are different to \eqref{eq:torsion} and contain zeroth order variations from $u_\mu$. But the current with coefficient equal to the charge susceptibility $\frac{d\rho}{d\mu^T_{5}}$ arises solely from torsion in the absence of other sources. See also \cite{ManesEtAl21}.

Of course, the fact that variation of $\delta \mu_{5} \sim \delta A^t_{5}$ is related to torsion follows from \eqref{eq:torsion} and this is the reason that the transport coefficient $\sigma_5$  with chiral anomaly U(1) contribution arises. This was merely due to a convenient way to compute the response: The background variation of $J^t_5$ for $\delta A^t_5\sim \mu_5$ with gauge fields is physically distinct from the (con)torsionful variation $\delta \omega^a_{\ b} \sim \delta T^a$. In the former, the background has no torsion, just a source of $A_{5}$, and, in the latter non-zero torsion is the only non-vanishing source. In particular, there is no vorticity $du =0$, since $\delta u_a = \delta e^a =0$. From the computation of CVE, the relation of the two responses is correct at the level of the linear response. The gravitational anomaly term $\propto T^2$ is ``explained" by the contorsionful spin-connection $\delta \hat{\omega}$. When both $\mu, \mu_5$ is non-zero, the hydrodynamics for a consistent (gauge invariant) theory need to be solved in addition, but the problematic vector-like terms are distinct from those above.

ii) Now the second independent linear response variation with torsion. Here we take the one-form $e^0= e^0_{\mu} dx^{\mu}$. In general, $e^a$ (or $e_a$) are varied with everything else \emph{fixed}; in particular $\hat{\omega}_{\mu} = 0$ and constant $u^a = (1,0,0,0)$.

These conditions set $\delta T^0 = d e^0 \neq 0$. Now $u = u_a e^a = u_0 e^0$. Then, following \cite{Zubkov2018, FerreirosLandsteiner20},
\begin{align}
\star J_5 &= c_v u \wedge du = c_v (u_0 e^0) \wedge d(u_0 e^0) \\
&= c_v u_0^2 e^0 \wedge de^0 = c_v e_0 \wedge d e^0 = c_v e_0 \wedge \delta T^0. \label{eq:CVE}
\end{align}
Again, one should assign this term exclusively to the background vierbein; it is linear in torsion and variation $\delta u$ changes as function of $\delta e^0_i$ only. This is the result originally by Khaidukov and Zubkov \cite{Zubkov2018}. The consistent choice is $\delta u_a=0$ in order to compute the exclusively torsional response probing non-zero $\delta T^0$; in the notation of \cite{FerreirosLandsteiner20}, the argument gives simply $c_{T\parallel} = c_v$ but not ``$c_{T\parallel} =0$". 

The above variation ii) is essentially the same argument about the connection of CVE to torsional current as the direct evaluation of the Kubo formula in \eqref{eq:torsion_Kubo} with non-trivial $e^0$. Combining the results i) and ii), we obtain the Kubo formula result \eqref{eq:torsion_Kubo}. On the other hand, if the take the variation of the metric in terms of $u_{\mu} = h_{t\mu}$ to linear order, it is impossible to maintain independently at the same time $\delta \hat{\omega}_\mu \neq 0$ and $\delta e^a_{\mu} =0$ ($\delta u_a \neq 0$) or vice versa, although the variations are related in linear response. Note that it is possible to consider the transformation to the rotating frame via the non-homogenous coordinate shift $\delta \Gamma^\lambda_{\mu\rho}$ \cite{FlachiFukushima18}, but this cancelled in $\delta \omega^a_{\mu b}$ up to local Lorentz rotations. Similarly, from the identification of $\vek{P}_i$ with $T^0_{\ i}$ or $T^{ti}$, the momentum cannot be independenty sourced with $e^a_i$ and $u_i$. The latter was exclusively used in \cite{FerreirosLandsteiner20}, see Eqs. \eqref{eq:u_variation}, \eqref{eq:tetrad_variations}, \eqref{eq:metric_variation}. On the other hand, the variation $\delta g_{\mu\nu} = h_{t\mu} = u_\mu$ with zero torsion produces the CVE with variations conjugate to momenta $T^{ti}$, Eq. \eqref{eq:metric_variation}.

This completes the detailed comparison of the Kubo formulas and chiral responses with and without torsion. To conclude, in our view, some of the conclusions of \cite{FerreirosLandsteiner20} (see also \cite{Chernodub2021}) regarding torsion are not correct. In particular, torsion leads to non-vanishing, independent currents with similar DC anomaly coefficients as for CVE. We have obtained the response in terms of explicit torsional currents, equivalent to results of previous work, utilizing the same linear response Kubo formulas. The only case when torsion does not contribute, is when it vanishes for the background (as is the case for the original Kubo formula computations). Again, it is a separate problem to ascribe the hydrodynamic degrees of freedom and constitutive relations with sources for a particular problem of chiral fermions and transport. Our results simply says that when torsion (i.e. tetrad and connection) is an independent hydrodynamic source variable, it activates chiral currents at finite chemical potential and temperature in relativistic systems. If only the metric enters, the response in terms of the CVE follows.

\section{Two gravitational anomalies in chiral Weyl condensates}\label{sec:cancellation}

Let us now discuss CVE and CTE in terms of a system where we believe both anomalies can be sourced independently: non-relativistic, chiral $p+ip$ paired condensate with Weyl quasiparticles at gap nodes. We note that while superficially both originate anomaly terms from either normal component or condensate vorticity, the arguments we presented above are of course independent of this property. For more discussion, see Sec. \ref{sec:Conclusions}.

\subsection{Relativistic low-energy theory}
We derive the low-energy theory for the normal component Bogoliubov fermions and the superfluid in a rotating vessel \cite{Volovik2003}. In rotating vessel, the normal component undergoes solid body rotation with $\vek{v}_n = \vek{\Omega} \times \vek{r}$, which in equilibrium is cancelled by a vortex lattice with spatially averaged $\langle \langle \vek{v}_s \rangle \rangle = \vek{v}_n$ over the unit cell.

Galilean transformations are $\vek{x}\to \vek{x}+\vek{v}t$. The superfluid free energy transforms as \cite{Volovik2003}
\begin{align}
f \to f[\vek{\rho}_s, \vek{l} ,\vek{v}_s] + \vek{g}_s\cdot \vek{v}
\end{align}
under small transformations $\vek{v}$. Here for the chiral $p$-wave $\vek{\Delta} = \Delta_0(\unitvek{m}+\im\unitvek{n})$ and $\unitvek{l} = \unitvec{m}\times\unitvec{n}$. The total mass current $\vek{g} = \vek{g}_{s} + \vek{g}_n$ transforms as $\vek{g} \to \vek{g}+\rho \vek{v}$, leading to $\vek{\rho}_n = \rho\id - \vek{\rho}_s$. The quasiparticles are governed by $[\im\doo_t -\mathcal{H}_{\rm BdG}(\vek{k})]\psi$ with the Bogoliubov-de Gennes Hamiltonian
\begin{align}
\mathcal{H}_{\rm BdG}(\vek{k}) = \left(\begin{matrix} \epsilon(\vek{k}) & \frac{1}{2}\{\vek{k},\vek{\Delta}\} \\ \frac{1}{2}\{\vek{k},\vek{\Delta}^*\} & -\epsilon(-\vek{k}) \end{matrix}\right) \label{eq:BdG_Hamiltonian}
\end{align}
where $\epsilon(\vek{k})$ is the normal state dispersion counted from the Fermi $k_F$ and the brackets $\{a,b\} =ab+ba$ preserve hermiticity with (weakly) coordinate dependent parameters. We assume equal spin-pairing and suppress spin indices. At the nodes $E_{\pm k_F\unitvek{l}} = 0$ are two Majorana-Weyl nodes (with spin degeneracy). The quasiparticles are momentum eigenstates transforming under Galilean boost as,
\begin{align}
E_{\vek{k}}(\vek{v}) = E_{\vek{k}} + \vek{k}\cdot \vek{v}.
\end{align}
More generally, the normal state fermions transforms $\Psi(\vek{x},t) \to e^{\im m \vek{v}\cdot \vek{x}}\Psi(\vek{x}+\vek{v}t,t)$, to lowest order in $v$, so the Bogoliubov fermions transform as $\psi(\vek{x},t) \to e^{\im m (\vek{v}\cdot \vek{x}-\mu t)\tau^0}\psi(\vek{x}-\vek{v}t,t)$. We generalize this to slowly-varying $\vek{\Delta}(\vek{x},t)$ and local transformations $\mu,\vek{v}$ in the gradient expansion, leading to the linearized $\im\doo_t -\mathcal{H}_{\rm BdG}$ close to $k_F\unitvek{l}$ as
\begin{align}
\tau^a e^\mu_aD_\mu \psi = \tau^a e_a^{\mu} [\doo_\mu - \im\hat{\omega}_{\mu}]\psi = 0, \label{eq:Linear_action}
\end{align}
where we redefined $\psi \to e^{-1/2} \psi$ and $\hat{\omega}_{\mu} = \frac{1}{2} \omega_{\mu}^{ab} \tau_{ab}=\omega^{12}_{\mu}[\tau_1,\tau_2]/2 = m(-\mu, \vek{v})\tau^3$. The (torsionful) BdG spin-connection is a Galilean boost connection with respect to the Newtonian time, which for the BdG quasiparticles acquires the form of a relativistic spin-1/2 connection in the $12$-plane. This spin-connection coincides with the combined $p+ip$ gauge symmetry U(1)$_{L_3-N/2}$ \cite{Nissinen20}. We note that using the anisotropic Newton-Cartan data \cite{Laurila20}, it can be equivalently written as the sum of a mass-current gauge field and a $\unitvek{l}$-orthogonal Christoffel connection (singular in the presence of vortices). The spin-connection is not needed here apart from the natural emergence of non-zero torsion along $\unitvek{l}$-direction \cite{Nissinen20, Copetti20, Laurila20}. 

In summary, the (inverse) tetrads from \eqref{eq:BdG_Hamiltonian} and \eqref{eq:Linear_action} are the linear expansion coefficients 
\begin{align}
e_0^{\mu} &= (1, -\vek{v}),\quad e_1^{\mu} = (0,c_{\perp}\unitvek{m}), \nonumber\\
e_2^{\mu} &= (0, c_{\perp}\unitvek{n}), \quad e_3^{\mu} = (0, c_{\parallel}\unitvek{l}), \label{eq:SF_tetrad}
\end{align}
where $c_{\perp} = \Delta_0/k_F$ and $c_{\parallel} = v_F$.  Inverting this, we obtain
\begin{align}
e_0^{\mu} &= (1, 0);\quad e_1^{\mu} = c_{\perp}(\unitvec{m}\cdot \vek{v},\unitvek{m}), \nonumber\\
e_2^{\mu} &= c_{\perp}(\unitvec{n}\cdot \vek{v},\unitvek{n}), \quad e_3^{\mu} = c_{\parallel}(\unitvek{l}\cdot \vek{v},\unitvek{l}) \label{eq:co-tetrad}.
\end{align}
There is the preferred (superfluid)frame where the normal component quasiparticles are at rest $\vek{v}_n = 0$, i.e. $\vek{v}=-\vek{v}_s$. The metric is secondary, and determines the shape of the linear Weyl dispersion \eqref{eq:Linear_action}.

\subsection{Cancellation of vortical and torsional anomaly currents}

In rotating chiral Wey superfluid/superconductor (e.g. $^3$He-A) with two components, normal and superfluid:
\begin{equation}
\vek{v}_n= \vek{u}= \boldsymbol\Omega \times \vek{r}, \quad \boldsymbol\Omega=\frac{1}{2} \boldsymbol\omega, 
\label{two_velocities}
\end{equation}
This produces the normal component CVE current $\vek{J}_{5} \propto \vek{\Omega}$ from the CVE \cite{VolovikVilenkin99}, via the Greens function identity \cite{Vilenkin79, FlachiFukushima18}
\begin{align}
G_{\vek{v}_n}(\vek{x},\vek{x}',k_0) = e^{-\frac{\im}{2}\Omega_3\Sigma_3 \doo_{k_0}}G_{\vek{v}_n=0}(\vek{x},\vek{x}',k_0)
\end{align}
where $k_0$ is the frequency, $\Sigma_3$ the spin-1/2 rotation matrix along $\unitvek{l}$ and we have specialized to $\vek{\Omega} = \Omega_3 \unitvek{l}$, so that the system is effectively relativistic to linear order in $\vek{k}$. 

On the other hand, from the BdG Hamiltonian to the linear order in $\vek{v}_s$, following \cite{Nissinen20}, the geometry \eqref{eq:SF_tetrad} induced by the superfluid background is torsionful. We set $\mu = \mu_s=0$, $\vek{v}=-\vek{v}_s$ with $\doo_{\mu}\unitvek{l} = \doo_{\mu}\unitvek{m} = \doo_{\mu}\unitvek{n} = 0$ and obtain the spatial torsion
\begin{align}
(\star e^a \wedge T_a)^k &= \frac{1}{c_\perp^2} \epsilon^{kijt}e^m_{i}\doo_j e^m_t + O(\vek{v}_s^2) \nonumber \\
&= -\frac{1}{c_{\perp}^2}\epsilon^{kij}\doo_i v_{s_j}, \quad m=1,2.
\end{align}
where the terms $\unitvek{l}\cdot \vek{v}_s = 0$ for the assumed vortex lattice configuration, see the discussion below in Sec. \ref{sec:Conclusions}. Note that this term is subleading in gradients compared to \cite{Nissinen20} but needs to be included, since $\nabla \times \vek{v}_n = \vek{\Omega}$ is non-zero.  

The combination of two gravitational anomalies gives CTE + CVE:
\begin{align}
J^\mu_5 &= \frac{T^2}{12}\epsilon^{\mu\nu\lambda\rho} \left[\frac{1}{2}\eta_{ab}e^a_{\nu}T^b_{\lambda\rho} + u_{\nu}\doo_{\lambda}u_{\rho}\right] \\
&= \frac{T^2}{12c_\perp^2} \left(\nabla \times \vek{v}_s -  2 \boldsymbol\Omega \right) . \nonumber
\label{combined}
\end{align}

In the equilibrium state in the rotating cryostat, the vortex lattice has spatially averaged vorticity $\langle \langle \nabla \times \vek{v}_s \rangle\rangle = 2\boldsymbol\Omega$, which corresponds to $\langle\langle \vek{v}_s\rangle\rangle= \vek{v}_n = \boldsymbol\Omega \times {\bf r}$ in equilibrium. That is why in the thermodynamic equilibrium the two anomalies cancel each other, $\langle \langle \vek{J}_5 \rangle\rangle=0$. The local currents exist  in the vortex lattice, $\nabla \times \vek{v}_s - \langle\langle\nabla \times  \vek{v}_s \rangle\rangle \neq 0$. But the total current along the rotation axis is zero. 

This demonstrates that a thermal version of the Bloch theorem \cite{Yamamoto15}, i.e. the absence of particle current in equilibrium, is applicable to the chiral vortical current (see e.g. Ref.\cite{Volovik2017} for the axial CVE current in $^3$He-A). Here the Bloch theorem is found as the consequence of the anomaly current cancellation in a chiral condensate with normal component chiral fermions.

\section{Conclusion and Outlook}\label{sec:Conclusions}

By extending the hydrodynamic chiral gravitational responses to non-zero torsion we computed the linear response Kubo formula with torsion and clarified the torsionful hydrodynamic variations, $u_\mu, g_{\mu\nu}$ vs. $e^a_{\mu}, \omega^a_{\mu b}$. We identify a simple reason why the CVE contributes at linear order with the gravitational $T^2$ term: it is directly related to the mixed torsional graviational anomaly with coefficient $t_{\chi} = d_{\chi}$. This complements the earlier arguments with thermodynamic variations and background sources formulated in terms of $T, \mu, u_i, g_{\mu\nu}$ utilizing the mixed gravitational $R^2$ anomaly but without torsion. In particular, recent work on anomalous (non-)relativistic transport with vierbeins \cite{Jain16, ManesEtAl21} did not take into account the presence of the Nieh-Yan anomaly terms, see however \cite{Copetti20}. The results we find with the relativistic Kubo formulas is in exact agreement with previous results, including the simple but physically compelling Landau level approach \cite{Laurila20, HuangEtAl20}.  

The relation of the CTE and CVE is reminiscent of the gravitional anomalies in general: for example, it is known that the non-conservation of the energy-momentum tensor is not independent of the Lorentz anomaly (i.e. antisymmetric $T^{ab}$) and that the anomaly can be shuffled from one to the other, as well as written in several equivalent forms \cite{AlvarezGaumeGinsparg85}. The choice where to place the anomaly should be made in terms of the independent degrees of freedom, the Lorentz and torsional anomalies being just manifestations of the presence of degrees of freedom independent from the metric. For hydrodynamic transport, the essential difference is whether the frame fields and connection enter the constitutive relations with sources independently from the metric. In the DC limit both conductivities are equal and constitute two parts: a chemical potential term $\mu^2$ induced by the chiral gauge anomaly and a $T^2$ term from the gravitational anomaly. In retrospect, this fits perfectly in to the different ways of understanding torsion in terms of a momentum dependent chiral gauge field corresponding to totally antisymmetric torsion. Finally, in order to discern between the two anomalies one can probe the finite frequency and momentum conductivities, which are different for the CTE and CVE, since the two independent contributions coincide only in the DC limit. On the other hand, it is known that the ``anomaly quantization" of the CVE, and therefore also the CTE, is only valid to the lowest non-trivial order $O(k^2)$ and that e.g. dynamical gauge fields contribute to the conductivity \cite{Golkar2015}.

We discussed the two anomalies in the example of non-relativistic chiral Weyl superfluids and superconductors, which in the low-energy approximation have chiral (Majorana-)Weyl fermions on anisotropic are Newton-Cartan geometries. Along the anisotropy direction and linear approximation, the geometry is effectively relativistic and the relativistic CVE and CTE results can be applied. The CVE is induced by the rotating normal component, whereas the torsionful tetrad arises from the superfluid component with $\vek{v}_s$. While CTE here superficially originates from condensate vorticity, we note that, by Mermin-Ho relations, $\vek{v}_s$ is related in hydrodynamics to the order parameter frame fields $e^a_i = \{\unitvek{m}, \unitvek{n}, \unitvec{l}\}$, of which only the $\unitvek{l}$ is a well-posed hydrodynamic variable in addition to, $\mu, T, \rho_s, \rho_n, \vek{v}_s, \vek{v}_n$ (spin is neglected). The perpendicular $\unitvek{m}, \unitvek{n}$-components contribute to $\vek{v}_s$ and represent gauge degrees of freedom in terms of the combined orbital-phase symmetry of the superfluid (superconductor) \cite{Volovik2003}. In the absence of relativistic systems with torsion, this comparison of the two anomalies with independent experimentally available sources is the best we can do. To see the relation of the two anomalies, we had to ``bias" $\vek{v}_s$ with $\vek{v}_n$ and go higher order in gradients as compared to \cite{Nissinen20}. In this case, the absence of chiral currents in equilibrium is obtained by the cancellation of the two anomaly currents, spatially averaged over the (unit cell) of the vortex lattice. To maintain the validity of the relativistic approximation with spin-1/2 chiral fermions, we considered the case where $\unitvek{l}$ is constant and along rotation axis $\vek{\Omega}$, corresponding to pure phase vortices \cite{SalomaaVolovikRMP}. 

In real rotating chiral superfluid $^3$He-A, the vortex lattice is created by quenching the critical rotation velocity whereby many different vortex lattices form quasi-equilibrium thermal states, with barriers exceeding 5 order of magnitudes compared to $T$. It is therefore expected that the anomaly can be ``transferred" between the $\unitvec{l}$-vector, $\vek{v}_s$ and $\vek{v}_n$ components of the superfluid and this choice is not gauge independent. It would be interesting to study this non-relativistic problem from the geometrical perspective. There is no particular hydrodynamical reason to restrict to low angular velocities, temperatures (compared to $T_c$), chemical potentials, where relativistically, to higher orders \cite{Vilenkin79, FlachiFukushima18}
\begin{align}
\vek{J}_5 = \left(\frac{\mu^2}{4\pi^2}+\frac{T^2}{12} + \frac{\Omega^2}{48\pi^2} - \frac{R}{96\pi^2}\right)\vek{\Omega} + \dots
\end{align}
with $R$ the scalar curvature in the plane perpendicular to $\vek{\Omega}$. Even better, the higher-order terms should be worked out in detail in terms of the full non-relativistic geometry and compared to the low-energy (relativistic) results and the regime of validity of the chiral transport with Weyl fermions, including torsion \cite{Imaki2020}. The ensuing effective hydrodynamic actions for the Goldstones should be worked out in detail.

On this note, we here for conceptual (and calculational \cite{Vilenkin79, FlachiFukushima18}) clarity interpreted the CVE exclusively from the rotation $\vek{\Omega}$ in the preferred frame with torsion. For consistency of the background, the CTE and CVE should be possible to understand solely in terms of the tetrad/metric, where the both CVE source $\vek{v}
_{n}$ and (torsionful) $\vek{v}_s$ tetrad directly contribute. In case of torsion, metric variations are superseded by those corresponding to the tetrad and connection. Moreover, Galilean invariance immediately suggests that $\vek{v}_s-\vek{v}_n$ should enter as the non-trivial component. Moreover, the non-relativistic extension of CTE could explain the $T=0$ anomalous angular momentum terms in the chiral $p+ip$ system \cite{VolovikVilenkin99}. Other non-relativistic systems where torsion is directly relevant are Weyl semimetals with dislocations: in the continuum approximation, the dislocation density arises from the torsion of globally non-trivial tetrad variation $\delta e^a_{\mu} = \doo_{\mu}u^a$, where $u^a$ is now the elastic displacement. In the case where $u^a$ is globally continuous, the non-trivial field strength needed for tetrads (or strain pseudo gauge fields) vanishes. In fact, $T^a_{\mu\nu}$ is equivalent to ``spatial vorticity" of the $u^a$ from dislocations and the findings of this paper can be applied with minimum alterations to axially twisted semimetals. What is missing is the spatial analogy to the static, equilibrium CVE and CTE is expected to play its role. In addition to twisting, the corresponding bias for dislocations could be the \emph{real} geometry/gravitational field, in the absence of rotation \cite{ZaanenEtAl21}. Nevertheless, see \cite{GaoEtAl21} for some recent considerations of the anomalous transport and CME \emph{without} dislocations (and therefore no effective torsion). 

{\bf Acknowledgements}. J.N. thanks Y. Ferreiros and K. Landsteiner for correspondence and explanations of their work. This work has been supported by the European Research Council (ERC) under the European Union's Horizon 2020 research and innovation programme (Grant Agreement No. 694248).  

\appendix
\section{Curvature conventions with torsion}
We discuss tetrads $g_{\mu\nu} = e^a_{\mu} e^b_{\nu}\eta_{ab}$ and connections $\hat{\Gamma}_\mu, \hat{\omega}_\mu$ that are metric compatible $\nabla_{\lambda} g_{\mu\nu} = \nabla_{\lambda} \eta_{ab} = 0$, 
\begin{align}
\nabla_\mu e^a_{\nu} &= \doo_{\mu}e^a_{\nu}-\Gamma^{\lambda}_{\mu\nu}e^a_{\lambda} + \omega^a_{\mu b} e^b_{\nu} = 0, \\
\Rightarrow \quad  \omega^a_{\mu b} &= e^a_{\lambda}\Gamma^\lambda_{\mu\nu} e^\nu_{b} + e^a_{\nu}\doo_{\mu} e^a_{\nu}.
\end{align}
We can form the well-defined tensor one forms in the tangent space $e^a = e^a_{\mu} dx^{\mu}$ and $\omega^{a}_{\ b} = \omega^a_{\mu b} dx^{\mu}$ that transform under local Lorentz rotations, the latter as a connection. The tetrad and connection are a priori independent quantities. and with field strength tensors that transform homogenously
\begin{align}
T^a = de^a + \omega^a_{\mu\nu} \wedge e^b =  \frac{1}{2}T^a_{\mu\nu} dx^{\mu} \wedge dx^{\nu} \\ 
R^a_{\ b}  = d\omega^a_{\ b} + \omega^a_{\ c }\wedge \omega^c_{\ b}  = \frac{1}{2} R^a_{\mu\nu b} dx^{\mu} \wedge dx^{\nu}
\end{align}
where in the coordinate basis
\begin{align}
T^\lambda_{\mu\nu} &= \Gamma^{\lambda}_{\mu\nu} - \Gamma^\lambda_{\nu\mu} \\
R^\lambda_{\mu\nu\rho} &= \doo_{\mu}\Gamma^\lambda_{\nu\rho} + \Gamma^{\lambda}_{\mu\tau}\Gamma^{\tau}_{\nu\rho} - (\mu \leftrightarrow \nu).
\end{align}
The curvature takes the usual form but depends on torsion through $\Gamma^{\lambda}_{\mu\nu}$. Without loss of generality, we can set
\begin{align}
\Gamma^\lambda_{\mu\nu} = \mathring{\Gamma}^\lambda_{\mu\nu} + K^\lambda_{\mu\nu}
\end{align}
where $K^{\lambda}_{\mu\nu} = \frac{1}{2}(T^{\lambda}_{\ \mu\nu} + T_{\nu\ \mu}^{\ \lambda} - T^{\ \ \lambda}_{\mu\nu})$ and $\mathring{\Gamma}^\lambda_{\mu\nu}[g_{\mu\nu}]$ is the symmetric Christoffel connection fully determined by the metric (or, equivalently, the tetrad $e^a_\mu$). The analogous formula holds for the metric compatible spin-connection $\hat{\omega} = \mathring{\omega}[e^a]+\hat{\omega}_{K}$.

For torsionful spacetimes the two tensors $T^a$ and $R^a_{\ b}$ charaterize the geometry. We can regard $e^a, \omega^a_{\ b}$ as independent but anothere option is to take $e^a$ and (con)torsion $T^a$ ($K^a_{\ b}$) as the independent variables. The corresponding variations are in \eqref{eq:torsion_variations}.

\section{Torsional Nieh-Yan form}
The Nieh-Yan form \cite{NiehYan1982a} is exact and given in terms of the 3-form with 1-form dual,
\begin{align}
e_a \wedge T^a  &= \frac{\eta_{ab}}{2} e^a_{\mu} T^b_{\nu\lambda} dx^\mu \wedge dx^{\nu} \wedge dx^\lambda, \\ 
\star e_a \wedge T^a &= \frac{\eta_{ab}}{2} \epsilon_{\rho}^{\ \mu\nu\lambda} e^a_{\mu} T^b_{\nu\lambda} dx^\rho
\end{align}
which are both (locally) Lorentz invariant tensors and therefore well-defined, in contrast to standard Chern-Simon like anomaly currents. The Bianchi identities with torsion imply that
\begin{align}
d(e_a \wedge T^a) = T^a \wedge T_a - e^a \wedge e^b \wedge R_{ab}.
\end{align}
This is an independent closed curvature invariant form that vanishes if and only if torsion is zero. See \cite{ChandiaZanelli97} for a discussion on the topological properties of the NY form and anomaly.

\section{Velocity perturbation}
Around flat space, we perturb $g_{\mu\nu}$ as $g_{\mu\nu} = \eta_{\mu\nu} + h_{\mu\nu}$. In the applications to CVE, $h_{ti} = u_i$, with all other components zero to linear order. The Christoffel connection $\mathring{\Gamma}_{\lambda\mu\nu} = g_{\lambda\rho} \mathring{\Gamma}^{\rho}_{\mu\nu}$ changes
as
\begin{align}
\delta\mathring{\Gamma}_{\lambda \mu \nu} &= \frac{1}{2}(\doo_{\mu}h_{\lambda\nu} + \doo_{\nu}h_{\mu\lambda} - \doo_{\lambda}h_{\mu\nu})\\
\delta\mathring{\Gamma}_{tij}& = \frac{1}{2}(\doo_i h_{tj}+\doo_j h_{ti})\\
\delta\mathring{\Gamma}_{itj}&= -\frac{1}{2}(\doo_{i}h_{tj}-\doo_{j}h_{ti})\\
\delta\mathring{\Gamma}_{itt} &= \doo_t h_{ti}.
\end{align}
The spin connection is
\begin{align}
\delta \mathring{\omega}_{a\mu b} & = e_{a\lambda} \delta \mathring{\Gamma}^\lambda_{\mu \nu} e^\nu_{b}+e_{a\nu}\doo_{\mu} e^\nu_b\\
& = \delta\mathring{\Gamma}_{a\mu b} + e_{a\nu}\doo_{\mu}e^\nu_b .
\end{align}
where $e_{a\nu}\doo_{\mu}e^{\nu}_b = \delta_{a0} \doo_{\mu}u_{m} \delta_{bm}$.
We get
\begin{align}
\mathring{\omega}_{0tm} &= \doo_{t}u_m, \\
\mathring{\omega}_{0im} &= \frac{1}{2}(\doo_i u_m - \doo_{m} u_i), \\
\mathring{\omega}_{mtn} &= \frac{1}{2}(\doo_m u_n - \doo_{n}u_m).
\end{align}
These lead to $T^a_{\mu\nu} = 0 + O(u^2)$ as expected. In the absence of torsion, the linear response to $u_{i} = h_{0i}$ is to be computed with the CVE Kubo formula.

\end{document}